\documentclass[a4paper]{jpconf}
\usepackage{graphicx}
\usepackage{multicol}

\newcommand{\mean}[1]{\langle#1\rangle}
\newcommand{\ud}[1]{{#1^{\dagger}}}

\begin{document}
\title{Photon correlations in both time and frequency}

\author{J. C. L\'opez Carre\~no, E. del Valle, F.P. Laussy}

\address{Departamento
de F\'isica Te\'orica de la Materia Condensada, Universidad
Aut\'onoma de Madrid, 28049 Madrid, Spain\\
Russian Quantum Center, Novaya 100, 143025 Skolkovo,
Moscow Region, Russia\\
Faculty of Science and Engineering, University of Wolverhampton, Wulfruna St, Wolverhampton WV1 1LY, UK}

%\ead{elena.delvalle.reboul@gmail.com}
\ead{f.laussy@wlv.ac.uk}

\section{Introduction}

Similarly to Satie's understanding of music as ``measuring
sound''% ~\cite{innes_book13a}
, Quantum Optics can be understood as the science of measuring photon
correlations. This led to one of the most recent revolutions
concerning our understanding of light, establishing among other things
that coherence is not a feature of monochromaticity but a quality of
photon correlators to factorize% ~\cite{glauber06a}
. Most of the field has concerned itself with time-resolved photon
correlations but to get a comprehensive picture, one should also
describe correlations in other variables (polarization, position,
etc.) One is particularly important in any dynamical system: energy.
Being a quantum mechanical problem, and since energy is $\hbar$ times
the frequency, this brings head on the problem of conjugate variables
and the uncertainty principle~\cite{arXiv_delvalle18a}.  Of particular
interest is Resonance Fluorescence (driving resonantly a two-level
system), whose spectral shape---the Mollow triplet% ~\cite{mollow69a}
---posed the question of peak-to-peak correlations in the early days
of the field~\cite{cohentannoudji77a}. Thanks to a new technique to
compute exactly frequency- and time-resolved photon
correlations~\cite{delvalle12a}, we have been able to provide the full
landscape of $N$-photon correlations from nontrivial
systems~\cite{gonzaleztudela13a}. This has been experimentally
measured~\cite{peiris15a} and found to be in excellent agreement with
the theoretical predictions.  Such an approach reveals a new class of
transitions, the \emph{leapfrog processes}, that involve $N$-photon
transitions over $N-1$ manifolds of excitation.  This makes for
strongly quantum correlated emission~\cite{sanchezmunoz14a}.
Intercepting (e.g., by filtering) such degenerate leapfrog transitions
allows to realize a regime of pure $N$-photon
emission~\cite{sanchezmunoz14a,sanchezmunoz18a}. The landscape of
correlation however extends far beyond the case of degenerate photons,
and a rich hyperspace of photon correlations exist at the
``\emph{photon-bundle}'' level~\cite{lopezcarreno17a}, where one can
correlate not simply photons but groups, or ``bundles'' of them.  This
should allow such a system to service a rich class of \emph{heralded
  $N$-photon sources}.  Such sources, if realized, would clearly have
strong implications for quantum spectroscopy (by exciting optical targets
with this new type of quantum
light)~\cite{lopezcarreno15a,lopezcarreno16a% ,lopezcarreno16b
}. In this text, written at the occasion of the METANANO 2018 in
Sochi, we present some illustrative and original results, highlighting
the temporal aspect on the one hand (for some fixed frequency) and the
frequency aspect on the other hand (for some fixed time, usually, the
case of coincidences, as this is the case which usually draws the
greatest attention).  While the formalism~\cite{delvalle12a} covers
these two aspects simultaneously, it is helpful and/or enlightening to
look at them separately. Since hastily reading observers may otherwise
get a wrong impression on the generality of the results, we also
present a case where correlations in \emph{both} time and frequency
are displayed. To also emphasize that the method is exact and see how
it compares with previous theories, that were not, we show explicit
calculations for both cases.

\section{Correlations in time}

\begin{figure*}[thbp]
  \centering
  \includegraphics[width=\linewidth]{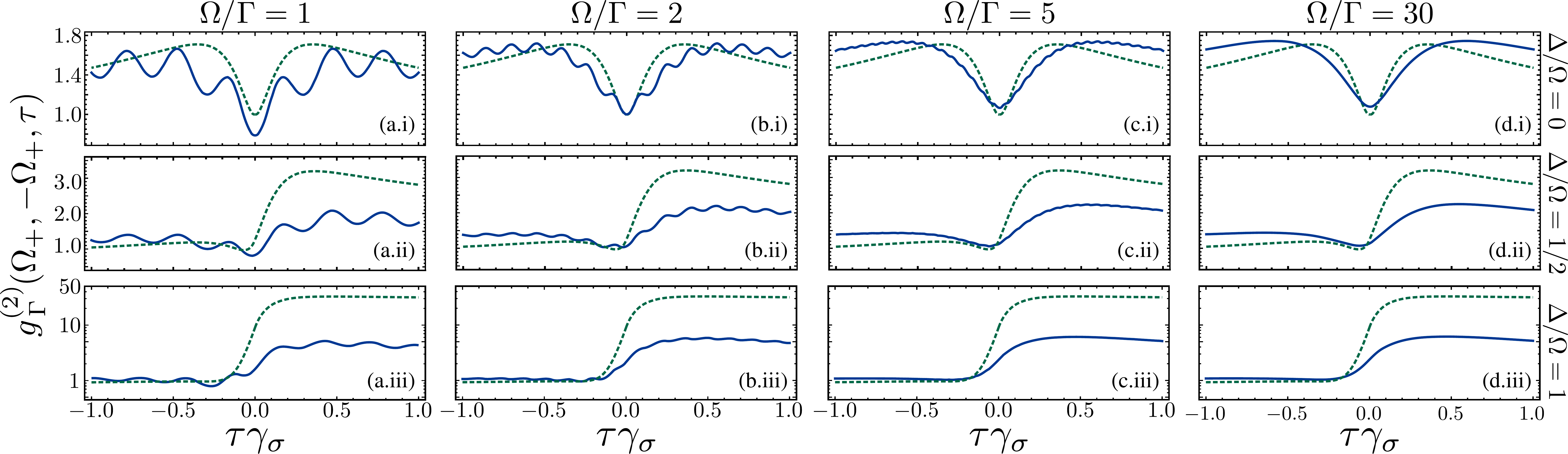}
  \caption{\small Comparison between the filtered correlations from
    the side peaks of the Mollow triplet, as obtained exactly through
    the sensing method~(solid blue) and approximately as in
    Ref.~\cite{schrama92a}~(dashed green). The driving
    intensity~$\Omega$ is varied from panels~(a) to~(d), and detuning
    between laser and emitter from panels~(i) to~(iii). Oscillations
    are lost in the approximation and the bunching peak associated to
    photon heralding is overestimated.  Parameters:
    $\Gamma/\gamma_\sigma=10$ ($\gamma_\sigma$ the emitter decay rate)
    and rest as shown.}
  \label{fig:Tue21Mar154846GMT2017}
\end{figure*}

Photon correlations are typically measured as a function of time-delay
between successive photons. With two photons, this yields the quantity
$g^{(2)}(\tau)$ (the so-called $2^{\mathrm{nd}}$ order correlation
function). The frequency-resolved version interposes filters in front
of the detectors and the quantity becomes
$g^{(2)}_{\Gamma}(\omega_1,\omega_2,\tau)$ (here with the same
frequency window~$\Gamma$ for both detectors, which needs not be the
case, see, e.g., Fig.~5 of Ref~\cite{gonzaleztudela15a}).  The formal
expression to compute was obtained exactly in the 80s, providing a
great result of photo-detection theory, but the actual computation
proved too complex and approximations had to be introduced. Typically,
following the fruitful dressed state picture, auxiliary states are
used and correlations between them computed as an approximation. We
will consider the case of correlations between the side peaks of the
Mollow triplet, and compare the results from
Ref.~\cite{schrama92a}(dashed green in
Fig.~\ref{fig:Tue21Mar154846GMT2017}), which rely on this
approximation, with our results (solid blue), which are numerically
exact. Overall, Physics is safe. The approximation is not exact and
can even fail qualitatively (most notably, it does not depend on
driving and misses oscillations at low driving) but the basic
behaviour is well captured and fairly accurate at high-driving and
resonance (the perfect agreement with experimental data is maybe less
satisfactory~\cite{schrama92a,ulhaq12a} but fitting the less accurate
model with free parameters can achieve that). The main limitation
however lies elsewhere.  Correlations are pinned to peaks emission in
this approximation, which motivates its form in the first place.
However, much more interesting correlations are to be found
\emph{away} from the peaks.
\begin{figure}[t]
  \centering
  \includegraphics[width=.85\linewidth]{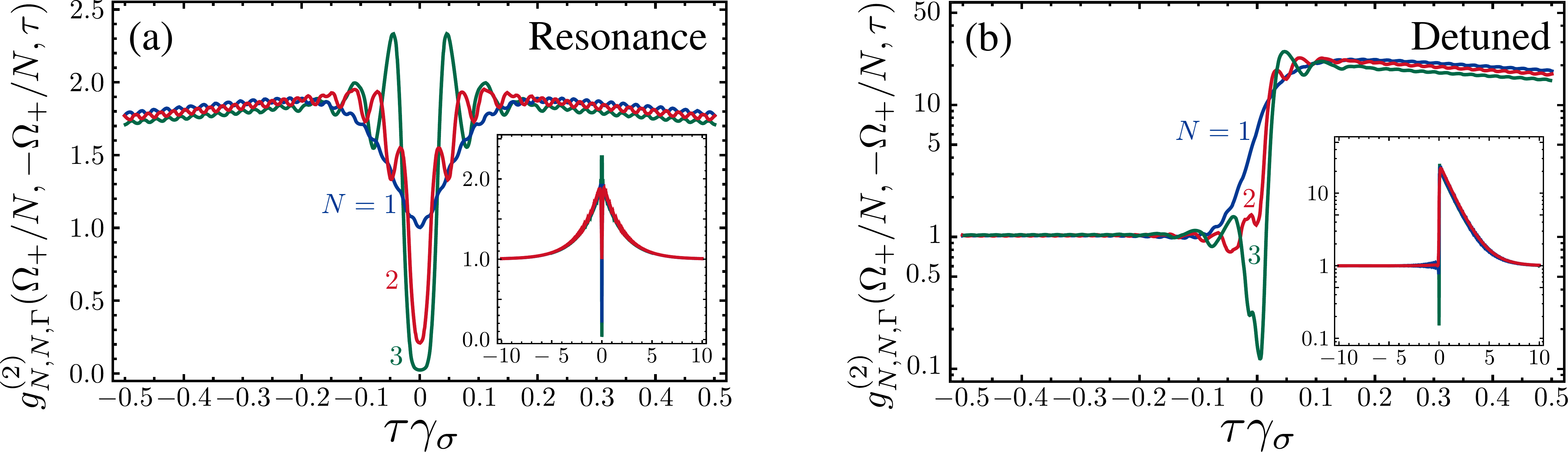}
  \caption{\small Two $N$-photon time cross-correlations at
    frequencies~$\pm\Omega_{+}/N$ (on both sides of the central peak of
    the Mollow triplet with $\Omega_+$ its splitting) for $N=1$~(blue
    lines), $2$~(red lines) and $3$~(green lines). (a)~Resonant
    excitation and (b)~detuned excitation (insets shown longer
    times). The detuning in~(b) is
    $\omega_\sigma=200\gamma_\sigma$. For all
    cases~$\Omega_+=300\gamma_\sigma$ and~$\Gamma=40\gamma_\sigma$,
    which maximizes the superpoissonian peak.}
  \label{fig:WedMar15134215GMT2017}
\end{figure}
Figure~\ref{fig:WedMar15134215GMT2017} shows $N$-photon correlations
at the frequencies $\pm\Omega_+/N$ for~$N=1$ (in blue,
cross-correlations of two single-photons from the side peaks, as in
Fig.~\ref{fig:Tue21Mar154846GMT2017}), $N=2$ (in red,
cross-correlations of two two-photon states \emph{halfway between the
  central and side peaks}) and $N=3$ (cross-correlations of two
three-photon states at one-third of the frequency between the central
and side peaks). Not only do we now access frequencies previously out
of reach, but we also consider correlations between \emph{bundles}
($N$-photon states), introducing the quantity
$g_{{n_1},{n_2},\Gamma}^{(2)}(\omega_1,\omega_2) \equiv {\mean{
    {:}\Pi_{\mu=1}^{2}\ud{a}^{n_\mu} (\omega_\mu)
    a^{n_\mu}(\omega_\mu){:}} }/(\Pi_{\mu=1}^{2}
\mean{\ud{a}^{n_\mu}(\omega_\mu) a^{n_\mu}(\omega_\mu) })$ (with
$a(\omega)$ the annihilation operator of a photon of
frequency~$\omega$ and~{:} normal ordering) that correlates a
$n_1$-photon bundle of frequency~$\omega_1$ with a~$n_2$-photon bundle
of frequency~$\omega_2$ both in frequency windows of width~$\Gamma$
(this can be generalized to different frequency
windows~\cite{lopezcarreno17a}). While single-photon correlations
become not particularly noteworthy, we see here that the same
phenomenology is transported to photon bundles. One can indeed
recognize in the insets of Fig.~\ref{fig:WedMar15134215GMT2017}(b) the
familiar profiles of photon heralding, but at the~$N$-photon
level. These results are thus not only exact, but also unique.

\section{Correlations in frequency}

\begin{figure*}[t]
  \centering
  \includegraphics[width=0.9\linewidth]{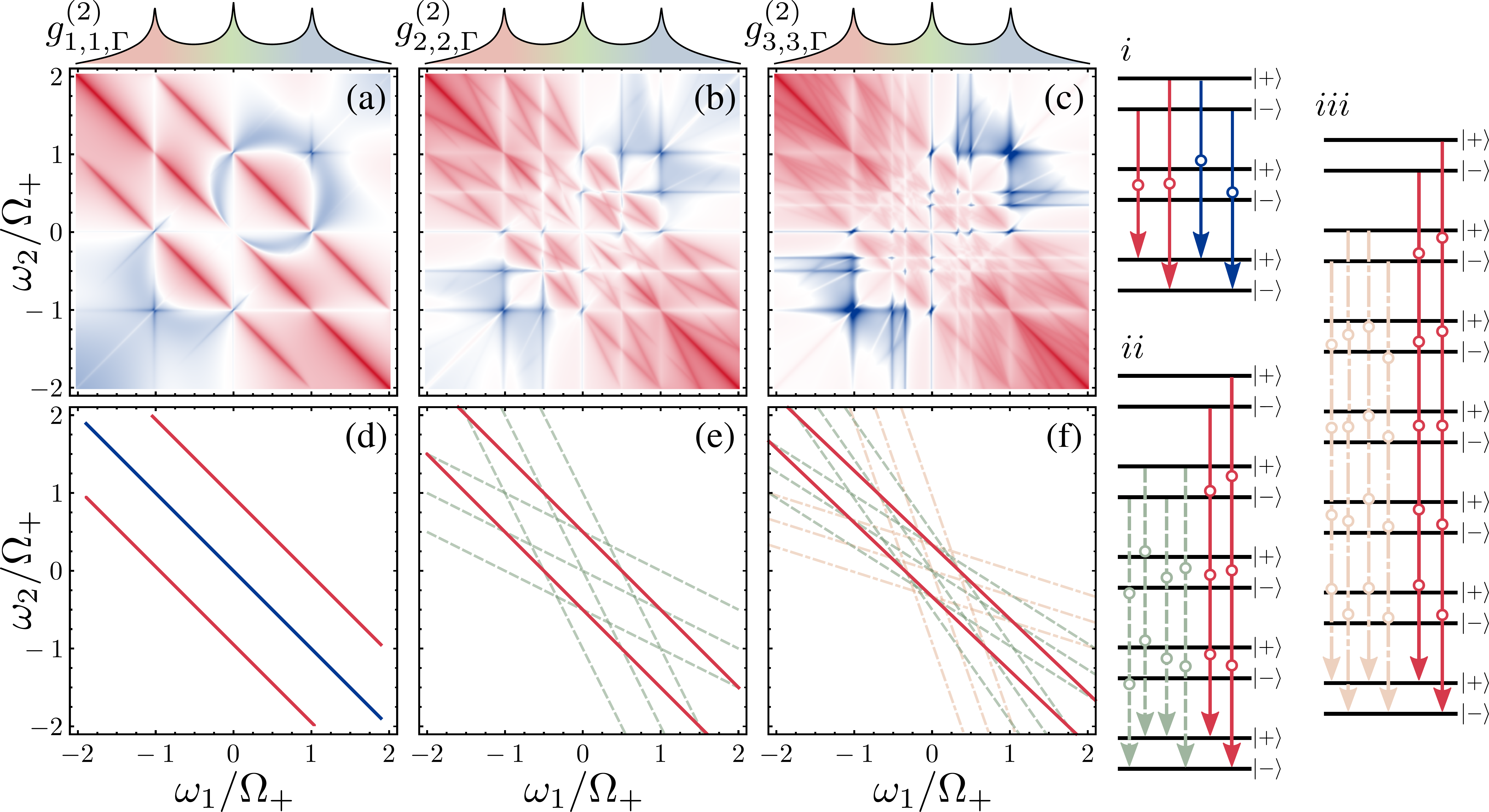}
  \caption{\small Landscape of two $N$-photon bundle correlations for
    (a) $N=1$, (b) $N=2$ and (c)~$N=3$, with red corresponding to
    bunching, blue antibunching and white no correlations. The
    structure, sketched in panels~(d)--(f), comes from leapfrog
    transitions, shown in insets $i$--$iii$ with corresponding
    colors. The structure in (d) is also present in~(b) and~(c) and
    that in~(e) is also present in~(c) (and so on to higher orders).
    Parameters: $\Omega_+=300\gamma_\sigma$, laser-emitter
    detuning~$\omega_\sigma = 200\gamma_\sigma$ and filtering window
    $\Gamma=5\gamma_\sigma$.}
  \label{fig:WedMar15143809GMT2017}
\end{figure*}

The correlation spectrum in frequency is probably the most conceptually
striking, as it brings a new dimension to the problem. It turns a
number, $g^{(2)}(0)$, into a picture, as shown in
Fig.~\ref{fig:WedMar15143809GMT2017}. Panel~(a) correlates two photons
of arbitrary frequencies, and is the case already measured
experimentally~\cite{peiris15a}. The antidiagonal red lines are
``leapfrog processes'' and show regions where the system tends to emit
photons in pairs.  Since we have amply discussed two-photon spectra in
previous works~\cite{gonzaleztudela13a}, we turn directly to the even
greater family of two $N$-photon bundles correlation
spectra. Panels~(b) and~(c) show correlations between two-photon and
three-photon bundles, respectively, here at resonance. In each case,
the structure of the correlation spectrum can be easily understood
from leapfrog transitions, shown in insets $i$--$iii$ along with, in
panels~(d)--(f), the traces they imprint in the correlation spectra,
with corresponding color codes. Note how case~$iii$, for instance,
involves six photons, in a plethora of photon emissions of various
orders, leading to a crowded landscape with myriads of
leapfrogs. Nevertheless, these are all clearly resolved
in~$g^{(2)}_{3,3}$ and understood in simple terms. Such correlated
emission can power devices.

\section{Correlations in both time and frequency}

\begin{figure}[!htb]
  \begin{minipage}{.45\linewidth}
    \centering
    \includegraphics[width=\linewidth]{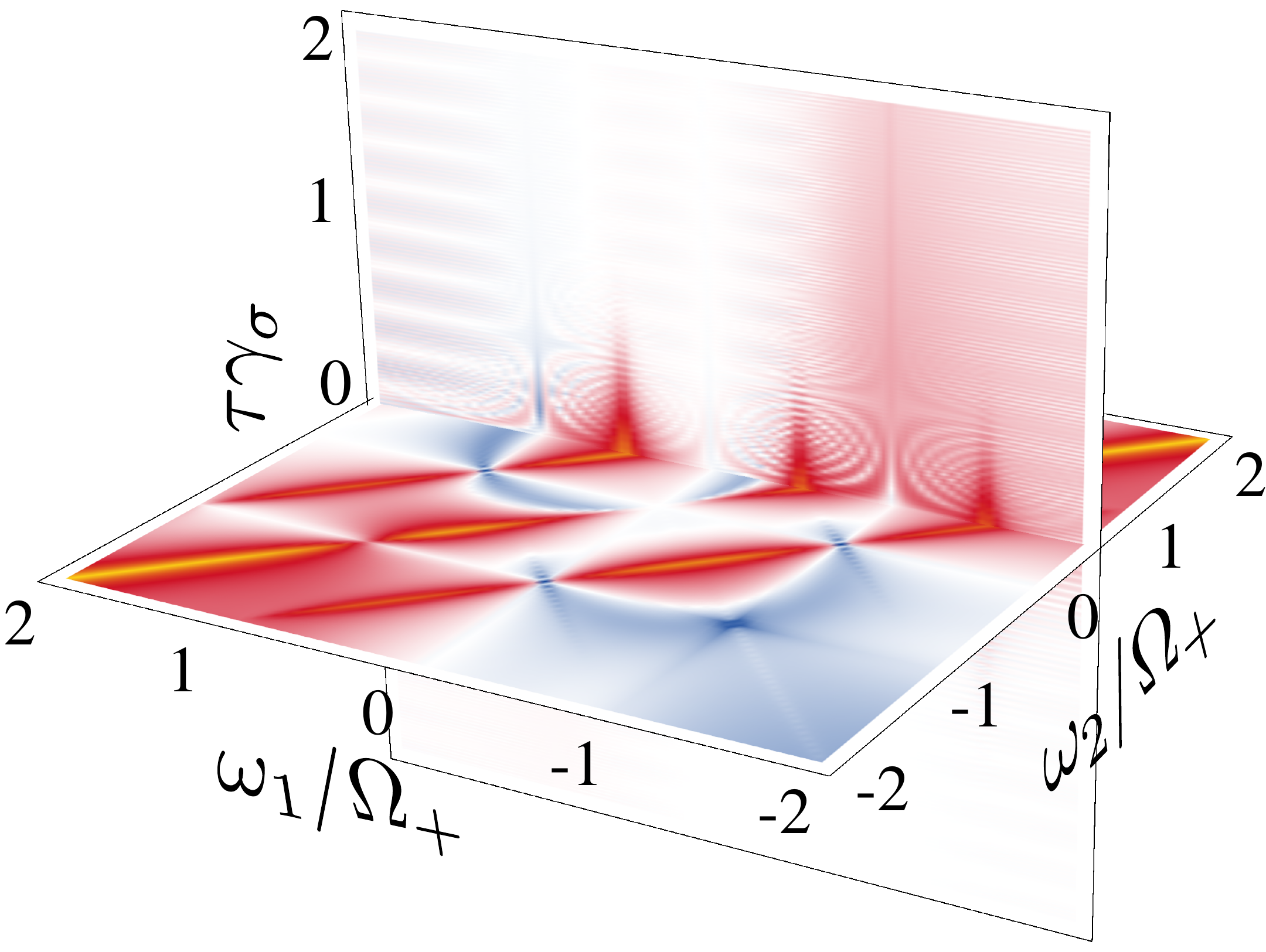}
    \label{fig:Fri23Feb180807GMT2018}
  \end{minipage}%
  \qquad
  \begin{minipage}{.45\textwidth}
    \vskip1cm\caption{\small Two-photon
      correlations~$g_\Gamma^{(2)}(\omega_1,\omega_2,\tau)$ from the
      Mollow triplet with a cut of time correlations at
      $\omega_2=\Omega_+/2$, showing how leapfrog processes imprint
      strong correlations in time along with the peaks emission. Of
      course the cut could be taken anywhere in the two-photon
      spectrum. Parameters: $\Gamma=5\gamma_\sigma$,
      $\Omega = 87 \gamma_\sigma$,
      $\Omega_+ = 174 \gamma_\sigma$ and driving at resonance.}
  \end{minipage}
\end{figure}

At this stage, it is clear that we have jointly considered
correlations in both time and frequency. We will let an image speak a
thousand words, with Fig.~\ref{fig:Fri23Feb180807GMT2018} showing
photon correlations simultaneously in time and frequency, in a 3D
space (that we have projected on two planes for clarity).  Heisenberg
uncertainties are satisfied through~$\Gamma$. There is no numerical
difficulty in getting these results, which have all been obtained with
a high-level programming language on a middle-end personal laptop in
time of the order of a few minutes. The only difficulty that arises is
not in obtaining the data in the first place, but in processing and
displaying it, as this starts to provide a fairly comprehensive
picture of a nontrivial quantum mechanical system.

% \begin{figure*}[thbp]
%   \centering
%   \includegraphics[width=.5\linewidth]{./figures/3D.pdf}
%   \caption{(Color online). }

% \end{figure*}

\section{Conclusions}

We have summarized the main results that follow from our theory of
frequency- and time-resolved photon correlations~\cite{delvalle12a},
which provides exact computations in nontrivial systems. We
illustrated the theory with original results as interesting cases can
be picked up from the endless configurations that abound in any
quantum optical system. In particular, we have compared our exact
results with previous approximations, upgraded correlations to the
case of bundles as otherwise no clear physical picture emerge,
highlighted the interest of measuring away from the peaks in rich
landscapes of correlations and emphasized the joint time and frequency
aspect of our theory. Such results can power a new class of optical
devices, out of which we foresee heralded $N$-photon emitters as the
most decisive breakthrough for today's photonics.

\section*{References}

% \begin{multicols}
%\bibliographystyle{naturemag}
\bibliographystyle{Science}
\bibliography{Sci,Books,arXiv} 

\begin{thebibliography}{10}

\bibitem{arXiv_delvalle18a}
E.~del Valle, J.~C.~L. Carre{\~n}o, F.~P. Laussy, {\it arXiv:1802.04540\/}
  (2018).

\bibitem{cohentannoudji77a}
C.~N. Cohen-Tannoudji, S.~Reynaud, {\it J. Phys. B.: At. Mol. Phys.\/} {\bf
  10}, 345 (1977).

\bibitem{delvalle12a}
E.~del Valle, {\it et~al.\/}, {\it Phys. Rev. Lett.\/} {\bf 109}, 183601
  (2012).

\bibitem{gonzaleztudela13a}
A.~Gonz\'alez-Tudela, {\it et~al.\/}, {\it New J. Phys.\/} {\bf 15}, 033036
  (2013).

\bibitem{peiris15a}
M.~Peiris, {\it et~al.\/}, {\it Phys. Rev. B\/} {\bf 91}, 195125 (2015).

\bibitem{sanchezmunoz14a}
C.~{S\'anchez Mu\~noz}, {\it et~al.\/}, {\it Nat. Photon.\/} {\bf 8}, 550
  (2014).

\bibitem{sanchezmunoz18a}
C.~{S\'anchez Mu\~noz}, {\it et~al.\/}, {\it Optica\/} {\bf 5}, 14 (2018).

\bibitem{lopezcarreno17a}
J.~C. {L\'opez Carre{\~n}o}, E.~del Valle, F.~P. Laussy, {\it Laser Photon.
  Rev.\/} {\bf 11}, 201700090 (2017).

\bibitem{lopezcarreno15a}
J.~C. {L\'opez Carre{\~n}o}, {\it et~al.\/}, {\it Phys. Rev. Lett.\/} {\bf
  115}, 196402 (2015).

\bibitem{lopezcarreno16a}
J.~C. {L\'opez Carre{\~n}o}, F.~P. Laussy, {\it Phys. Rev. A\/} {\bf 94},
  063825 (2016).

\bibitem{schrama92a}
C.~A. Schrama, {\it et~al.\/}, {\it Phys. Rev. A\/} {\bf 45}, 8045 (1992).

\bibitem{gonzaleztudela15a}
A.~Gonz\'alez-Tudela, E.~del Valle, F.~P. Laussy, {\it Phys. Rev. A\/} {\bf
  91}, 043807 (2015).

\bibitem{ulhaq12a}
A.~Ulhaq, {\it et~al.\/}, {\it Nat. Photon.\/} {\bf 6}, 238 (2012).

\end{thebibliography}
% \end{multicols}
\end{document}